\documentclass[twoside]{dis09}
\usepackage[latin1]{inputenc}
\usepackage[dvips]{graphicx,epsfig,color}
\usepackage{wrapfig,rotating}
\usepackage{amssymb,amsmath,array}

\begin{document}
\title{Forward Physics Capabilities of CMS with the CASTOR and ZDC detectors}

%***********************************************************************
% AUTHORS INFORMATION AREA
%***********************************************************************
\author{Roland Beno\^it
%
% Optional short acknowledgment: remove next line if non-needed
%\thanks{This is an optional funding source acknowledgment.}
%
% DO NOT MODIFY THE FOLLOWING '\vspace' ARGUMENT
\vspace{.3cm}\\
%
% Addresses and institutions (remove "1- " in case of a single institution)
Universiteit Antwerpen - Physics Department \\ 
Groenenborgerlaan 171, Antwerpen - Belgium
%
% Remove the next three lines in case of a single institution
%\vspace{.1cm}\\
%2- School of Second Author - Dept of Second Author \\
%Address of Second Author's school - Country of Second Author's school\\
}
%***********************************************************************
% END OF AUTHORS INFORMATION AREA
%***********************************************************************

\maketitle

\begin{abstract}
%Place your abstract here. It should not exceed 100 words.
The two calorimeters CASTOR and ZDCs enhance the hermeticity of the CMS detector at the LHC
by extending the rapidity coverage in the forward region. After having described these detectors, 
their forward physics capabilities are presented. These latters include the study of parton shower, 
multiple parton interactions, diffraction and ultra high energy cosmic rays models. The processes 
to be measured to constrain these topics are multi-jet events with a forward jet, 
central-forward activity correlation, rapidity gaps and forward neutron production.   
\end{abstract}

\section{The forward calorimeters CASTOR and ZDCs}

\begin{wrapfigure}{r}{0.5\columnwidth}
\centerline{\includegraphics[width=0.45\columnwidth]{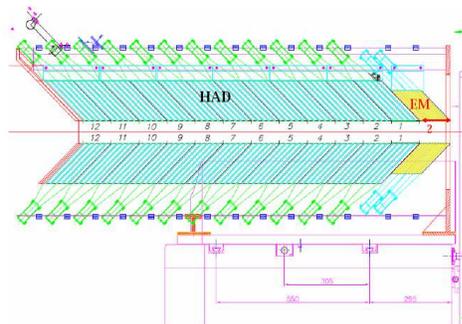}}
\caption{The CASTOR calorimeter.}\label{Fig:CASTOR}
\end{wrapfigure}

The CASTOR \cite{Aslanoglou:2007wv} calorimeter, illustrated on Figure \ref{Fig:CASTOR}, is located 14.37 m 
from the CMS interaction point and extends the forward rapidity coverage to the region $-6.6 < \eta < -5.2$.
CASTOR presents a sandwich structure of tungsten (W) absorber plates and quartz plates as active material. It has an 
octagonal cylinder shape with an inner radius of 3.7 cm, an outer radius of 14 cm and a total depth of 10.3 $\lambda_I$.
The collection of the signal is based on the production of \v{C}erenkov photons which are transmitted to 
photomultiplier tubes through aircore lightguides, the W and quartz plates being inclined by 45 \mbox{$^{\circ}$}
w.r.t. the beam axis to optimize the photon yield. The CASTOR calorimeter is divided in an electromagnetic section 
of 20.12 $X_0$ and an hadronic section of 9.5 $\lambda_I$. It has a 16-fold azimuthal segmentation in towers 
and a 14-fold longitudinal segmentation in sections, from which the 2 first are electromagnetic and the 12 remaining 
hadronic. It has no segmentation in $\eta$ and consists therefore 
in a total of 224 channels. Castor has just been submitted to final tests in beam and installed inside of the 
CMS cavern at respectively the beginning and the end of this month of June 2009. \\

\begin{wrapfigure}{r}{0.5\columnwidth}
\centerline{\includegraphics[width=0.40\columnwidth,clip=]{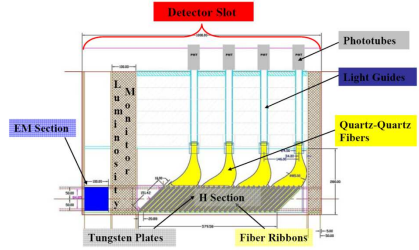}}
\caption{The ZDC calorimeter.}\label{Fig:ZDC}
\end{wrapfigure}

The Zero Degree Calorimeters \cite{Grachov:2006ke} (ZDCs), illustrated on Figure \ref{Fig:ZDC}, are located 140 m 
from the interaction point in both the forward and backward directions. They enable to measure neutral particles 
produced at $|\eta| > 8.1$ and possess a full acceptance to measure neutral energy flow in the region $|\eta| > 8.3$.
The ZDCs are also based on the production and detection of \v{C}erenkov photons to build the calorimeter signal 
and present as CASTOR a sandwich structure of W plates as absorber and quartz fibers as active material.      
Each of the ZDC calorimeters is divided in an electromagnetic section of 19 $X_0$ and an hadronic section 
of 5.6~$\lambda_I$. The electromagnetic section has a 5-fold horizontal segmentation to measure the pseudorapidity
of the forward energy deposits, while the hadronic section has a 4-fold longitudinal segmentation.   
The ZDCs were already integrated into CMS for the 2008 first LHC run.\\ 

\section{The CASTOR and ZDCs forward physics capabilities}

The theoretical description of $p \, p$ collisions at high energies involves the use of Matrix Elements (ME)
associated to the hard scattering and Parton Showers (PS) linking the scattered partons with the final-state hadrons. 
The ME can be computed exactly at a given fixed order of the perturbative QCD expansion
and for given values of the hard scale $Q^2$ and Bj\"orken variable $x$, while the PS takes into account
higher order contributions by resuming a subset of leading diagrams at each order. The subset of leading
diagrams considered depends on the values of $x$ and $Q^2$ and various models exist to describe
the PS evolution in the different phase-space regions. 

\subsection{Multi-jet events with a forward jet}

\begin{wrapfigure}{r}{0.5\columnwidth}
\vspace*{-0.8cm} 
\centerline{\includegraphics[width=0.45\columnwidth]{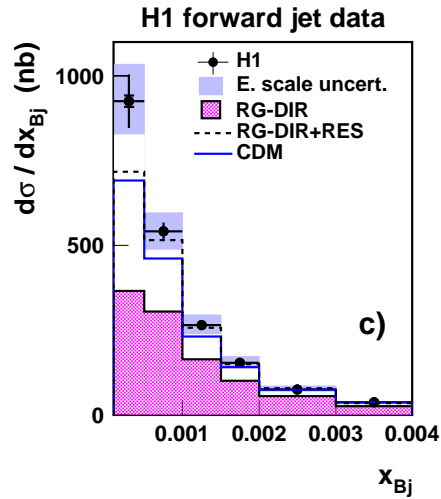}}
\caption{The hadron level cross section for forward jet production as a function of $x_{bj}$ 
compared to the QCD Monte Carlo models RAPGAP \cite{Jung:1993gf} and Ariadne \cite{Lonnblad:1992tz}.}
\label{Fig:FwdJetH1}
\end{wrapfigure}

The interest of forward physics to study the PS evolution comes from the fact that the differences 
between the predictions of the various models are more prominent in the forward region. 
The DGLAP \cite{DGLAP} evolution equations as implemented in PYTHIA \cite{Sjostrand:2006za}
and for which the gluon ladder is ordered in $k_t$ predict PS in which the softest emissions are the ones closest 
to the proton remnant direction, while the BFKL \cite{BFKL} evolution equations for which the gluon ladder is ordered 
in $x$ without any $k_t$ ordering or the Color Dipole Model (CDM) as implemented in Ariadne \cite{Lonnblad:1992tz} and 
in which the emissions are generated by color dipoles radiating independently predict arbitrarily large forward 
emissions as long as allowed by the kinematics.  The measurement of multi-jet events with a forward jet in CASTOR 
is therefore an important tool to distinguish between the various models and to study the PS dynamics beyond 
the standard direct DGLAP approximation. This is illustrated on Figure \ref{Fig:FwdJetH1} by the H1 results 
relative to the study of forward jet production in Deep Inelastic Scattering at HERA \cite{Aktas:2005up}. 
The data are compared to the QCD Monte Carlo models RAPGAP \cite{Jung:1993gf} and Ariadne \cite{Lonnblad:1992tz},
which respectively uses the DGLAP evolution equations and the Color Dipole Model to describe the Parton Shower. \\

The study of the distribution of the pseudorapidity separation between jets, $\Delta \eta$, in multi-jet events 
with a forward jet in CASTOR, would also enable to distinguish between the various PS models. By selecting events
with different $\Delta \eta$ topologies, one can indeed look at the breaking of the $k_t$ ordering at different
places along the gluon ladder. One can for example enhance the available phase-space in $x$ for BFKL-type radiations
by looking to events with a forward jet in CASTOR and a central dijet system \cite{Aktas:2005up}. \\

The CASTOR calorimeter can also be used to measure Mueller-Navelet dijet events \cite{MNdijet} in which a jet 
is detected in each of the forward directions. This process is characterized by the presence of two hard scales 
- the transverse momentum of the two jets -  which leads to a suppression of the emissions 
ordered in $k_t$ as described by the DGLAP equation. It is also characterized by a large
rapidity interval between the two jets of the dijet system,  which leads to an enhancement of the 
BFKL-type radiations. The measurement of Mueller-Navelet dijet events is therefore well suited
to access the dynamics beyond DGLAP, particularly through the study of the azimuthal
decorrelation $\Delta \phi$ between the two jets of the system \cite{Marquet:2007xx}. \\

%\begin{wrapfigure}{r}{0.5\columnwidth}
%\centerline{\includegraphics[width=0.40\columnwidth]{roland_benoit.fig4.eps}}
%\caption{Correlation between the azimuthal angle of the jet at generated level and CASTOR level.}
%\label{Fig:Knutsson1}
%\end{wrapfigure}
\begin{figure}[h!]
\centerline{\includegraphics[width=0.35\columnwidth]{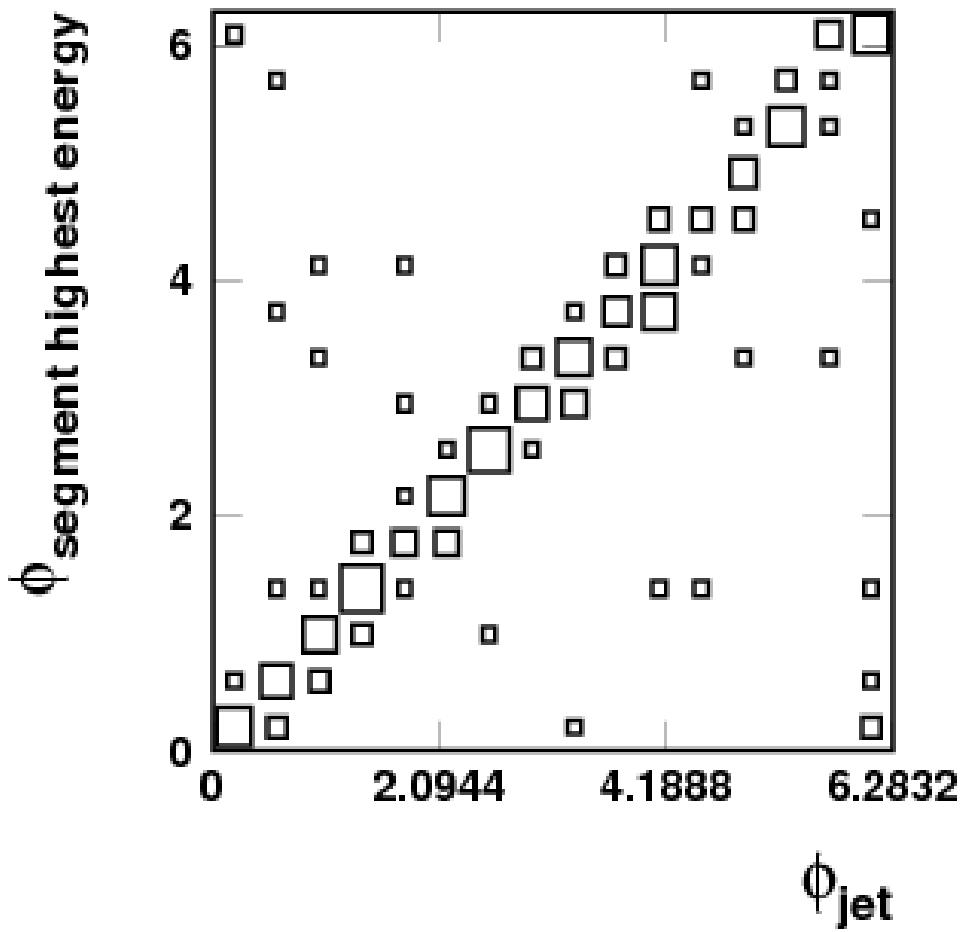}
\includegraphics[width=0.35\columnwidth]{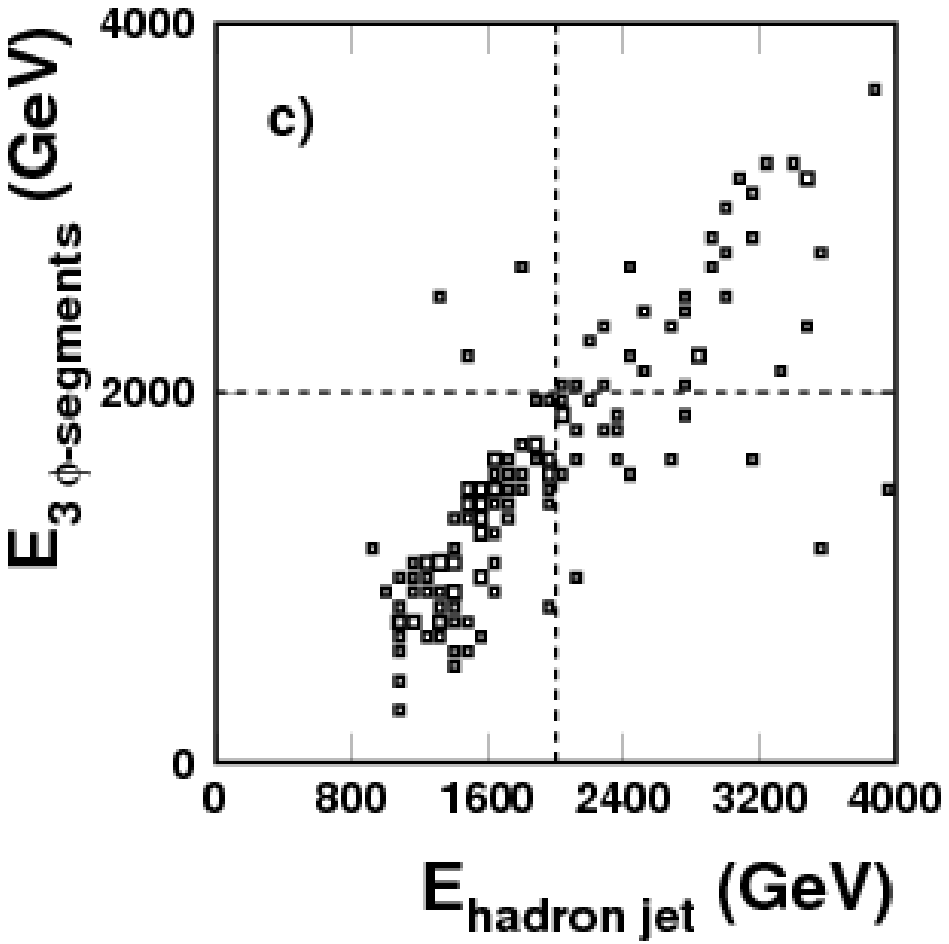}}
\caption{Azimuthal angle and energy of the jet at generated level and CASTOR level.}
\label{Fig:Knutsson}
\end{figure}
The feasibility to reconstruct forward jets in CASTOR has been studied by the Antwerpen and DESY groups.
The results presented here have been obtained by Albert Knutsson and correspond to a generator level study
based on Ariadne. The jets are defined at generator level according to the inclusive $k_t$ algorithm and 
at CASTOR level according to a simplified approach by considering 16 sectors in $\phi$ and summing the 
energy longitudinally in each of these sectors. The azimuthal angle of the jet at CASTOR level  is then given 
by the azimuthal angle of the highest energetic sector, while the energy of the jet is given 
by the energy of the highest energetic sector to which the energies of the two neighboring ones are added.
One has to emphasize that no detector effects are taken into account. The correlation 
between the energy and the azimuthal angle of the jet at generated level and CASTOR level
is shown on Figure \ref{Fig:Knutsson} and found to be reasonably good.  
%\begin{wrapfigure}{r}{0.5\columnwidth}
%\centerline{\includegraphics[width=0.40\columnwidth]{roland_benoit.fig5.eps}}
%\caption{Correlation between the energy of the jet at generated level and CASTOR level.}
%\label{Fig:Knutsson2}
%\end{wrapfigure}

\subsection{Underlying Events and Multiple Parton Interactions}

\begin{figure}[h!]
\centerline{\includegraphics[width=1.0\columnwidth,clip=]{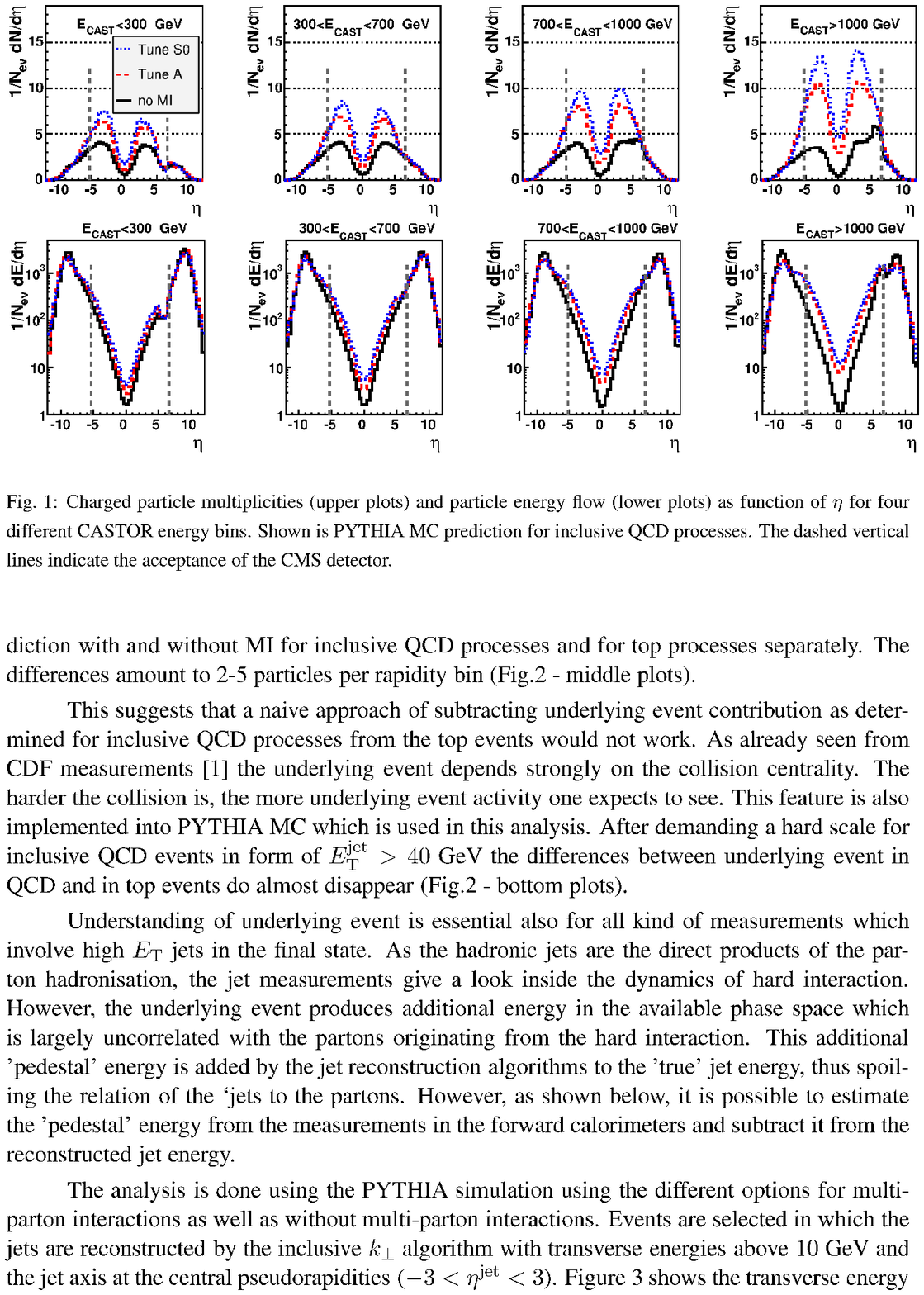}}
\caption{Charged particles multiplicities as a function of $\eta$ for different CASTOR energy deposits. 
Shown is PYTHIA prediction for inclusive QCD processes and several MPI tunes.}
\label{Fig:Correlation}
\end{figure}

The Underlying Event (UE), defined as everything except the hard scattering component of a collision, 
is an unavoidable background which has direct implications on measurements at the LHC.
It affects for example the jet reconstruction and the isolation cuts. The UE receives
contributions from the Initial and Final State Radiation (gluon emissions), from the Beam Remnants
(particles coming from the proton break-up) and from the Multiple Parton Interactions (MPI)
which correspond to additional softer parton scatterings. If from an experimental point of view, 
it is impossible to separate the UE from the hard component, the topological structure of a $p \, p$ collision 
however enables to define observables which are sensitive to the UE, like the average sum of charged particles 
transverse momentum in the transverse region w.r.t. the leading jet direction. The problem in the understanding 
of the UE at the LHC comes from the fact that different MPI models tuned to describe equally well Tevatron data 
for the above observable give very different predictions when extrapolated at the LHC energy \cite{UE}. The
MPI occurring between the spectator partons, they strongly affect the forward energy flow and the measurement 
of energy deposit in CASTOR could therefore be sensitive to the various MPI models. The MPI induced correlation 
between the activities in the forward and central regions could furthermore be studied by looking
at the charged particles multiplicities as a function of $\eta$ for different energy deposits in CASTOR.
Figure \ref{Fig:Correlation} illustrates such a study on the basis of a 
generator level analysis of inclusive QCD processes made with PYTHIA for several MPI tunes \cite{Bunyatyan}.
  
\subsection{Diffraction and rapidity gaps}

Diffractive $p \, p$ interactions, for which one or both protons stay intact, enable to study the perturbative QCD  
and the hadron structure through the measurement of the cross sections for diffractive jet, $Z$, $W$ or heavy quark 
production. Diffraction also enables to study MPI and soft rescattering through the extraction of the rapidity 
gap survival probability. Diffractive processes will be selected at CMS by rejecting events with a forward activity 
in the Hadron Forward (HF) and CASTOR calorimeters. The use of CASTOR could yield in particular to an improved
rejection of non-diffractive processes, while the ZDCs could be used to reduce the diffractive dissociation
background \cite{DiffW}.

\subsection{ZDCs $p \, p$ forward physics program}

The ZDCs being fast enough to give an answer at the level 1 trigger, they could be used online
to select the diffractive events and to reject the diffractive dissociation background.  
The ZDCs also plan to measure the correlation between the neutral forward energy flow and
the particle multiplicity in the central region. The measurement of the forward neutron production
by the ZDCs could lead in particular to a better constraint on the low $x$ part of the gluon pdf and 
the Ultra High Energy cosmic rays models \cite{d'Enterria:2008jk}. 
The ZDCs could finally monitor the luminosity of the beam
by looking at $p \, p$ bremsstrahlung events and forward neutron production.

\section{Conclusion}

The forward calorimeters CASTOR and ZDCs are planed to play an important role in the study
of forward physics with the CMS detector at the LHC. Their program, extending from forward jet
production to diffraction, could enable to measure a wide variety of observables.

\section{Acknowledgments}

I would like to thank my colleagues of the CMS Forward Physics Analysis Group for the comments,
discussions and for providing the results presented in these proceedings.

\section{Bibliography}

% ****************************************************************************
% BIBLIOGRAPHY AREA
% ****************************************************************************

\begin{footnotesize}

% IF YOU DO NOT USE BIBTEX, USE THE FOLLOWING SAMPLE SCHEME FOR THE REFERENCES
% ----------------------------------------------------------------------------

% ----------------------------------------------------------------------------

% IF YOU USE BIBTEX,
% - DELETE THE TEXT BETWEEN THE TWO ABOVE DASHED LINES
% - UNCOMMENT THE NEXT TWO LINES AND REPLACE 'Name_Of_Your_BibFile'

%\bibliographystyle{unsrt}
%\bibliography{Name_Of_Your_BibFile}

\begin{thebibliography}{99}
% Please replace the numbers for   contribId   and   sessionId
% in the following URL. You can get this information by going to 
% http://indico.cern.ch/confAuthorIndex.py?confId=24657
% and search for your contribution and click on the title
% Be aware: '&amp;' must be replaced by simple '&' as in example below

%\bibitem{url} Slides: \\ 
%\verb$http://indico.cern.ch/contributionDisplay.py?contribId=23&sessionId=22&confId=53294$

%------- replace following references 

\bibitem{Aslanoglou:2007wv}
X.~Aslanoglou {\it et al.},
%``Performance Studies of Prototype II for the CASTOR forward Calorimeter at the CMS Experiment,''
Eur.\ Phys.\ J.\  C {\bf 52} (2007) 495
[arXiv:0706.2641 [physics.ins-det]].
%%CITATION = EPHJA,C52,495;%%

\bibitem{Grachov:2006ke}
O.~A.~Grachov, M.~J.~Murray, A.~S.~Ayan, P.~Debbins, E.~Norbeck, Y.~Onel and D.~G.~d'Enterria [CMS Collaboration],
%``Status of zero degree calorimeter for CMS experiment,''
AIP Conf.\ Proc.\  {\bf 867} (2006) 258
[arXiv:nucl-ex/0608052].
%%CITATION = APCPC,867,258;%%

\bibitem{DGLAP} V.~Gribov and L.~Lipatov, Sov.\ J.\ Nucl.\ Phys.\ {\bf 15} (1972) 438 [Yad. Fiz. {\bf 15} (1972) 781];\\
               V.~Gribov and L.~Lipatov, Sov.\ J.\ Nucl.\ Phys.\ {\bf 15} (1972) 675 [Yad. Fiz. {\bf 15} (1972) 1218];\\
               Y.~Dokshitzer, Sov.\ Phys.\ JETP {\bf 46} (1977) 641 [Zh. Eksp. Teor. Fiz. {\bf 73} (1977) 1216];\\
               G.~Altarelli and G.~Parisi, Nucl.\ Phys.\ B {\bf 126} (1977) 298.

\bibitem{Sjostrand:2006za}
T.~Sjostrand, S.~Mrenna and P.~Skands,
%``PYTHIA 6.4 Physics and Manual,''
JHEP {\bf 0605} (2006) 026
[arXiv:hep-ph/0603175].
%%CITATION = JHEPA,0605,026;%%

\bibitem{BFKL} E.~A.~Kuraev, L.~N.~Lipatov and V.~S.~Fadin, Sov.\ Phys.\ JETP {\bf 45} (1977) 199 [Zh. Eksp.
    Teor. Fiz. {\bf 72} (1977) 377];\\
    I.~I.~Balitsky and L.~N.~Lipatov, Sov.\ J.\ Nucl.\ Phys.\ {\bf 28} (1978) 822 [Yad. Fiz. {\bf 28} (1978) 1597];\\
    L.~N.~Lipatov, Sov.\ Phys.\ JETP {\bf 63} (1986) 904 [Zh. Eksp. Teor. Fiz. {\bf 90} (1986) 1536].

\bibitem{Lonnblad:1992tz}
L.~Lonnblad,
%``Ariadne Version 4: A Program For Simulation Of QCD Cascades Implementing The Color Dipole Model,''
Comput.\ Phys.\ Commun.\  {\bf 71} (1992) 15.
%%CITATION = CPHCB,71,15;%%

\bibitem{Aktas:2005up}
A.~Aktas {\it et al.}  [H1 Collaboration],
%``Forward jet production in deep inelastic scattering at HERA,''
Eur.\ Phys.\ J.\  C {\bf 46} (2006) 27
[arXiv:hep-ex/0508055].
%%CITATION = EPHJA,C46,27;%%

\bibitem{Jung:1993gf}
H.~Jung,
%``Hard diffractive scattering in high-energy e p collisions and the Monte Carlo generator RAPGAP,''
Comput.\ Phys.\ Commun.\  {\bf 86} (1995) 147.
%%CITATION = CPHCB,86,147;%%

\bibitem{MNdijet}
A.~H.~Mueller and H.~Navelet, Nucl.\ Phys.\  {\bf B282} (1987) 727.

\bibitem{Marquet:2007xx}
S. Cerci, D. d'Enterria, ``DIS 2009, these proceedings'';\\
C.~Marquet and C.~Royon,
%``Azimuthal decorrelation of Mueller-Navelet jets at the Tevatron and the LHC,''
Phys.\ Rev.\  D {\bf 79} (2009) 034028
[arXiv:0704.3409 [hep-ph]]. 
%%CITATION = PHRVA,D79,034028;%%

\bibitem{UE}
C.~Buttar {\it et al.}, Physics at Tev Colliders 2005 - QCD, EW and Higgs Working Group: Summary Report, April 2006.

\bibitem{Bunyatyan}
A.~Bunyatyan and Z.~Rurikova, 
``Underlying Event Studies with CASTOR in the CMS Experiment,''
HERA-LHC proceedings.

\bibitem{DiffW}
A.~V.~Pereira, ``Single diffractive W production with CMS,'' CMS PAS DIF-07-001.

\bibitem{d'Enterria:2008jk}
D.~d'Enterria, R.~Engel, T.~McCauley and T.~Pierog,
%``Cosmic-ray Monte Carlo predictions for forward particle production in p-p,
%p-Pb, and Pb-Pb collisions at the LHC,''
arXiv:0806.0944 [astro-ph].
%%CITATION = ARXIV:0806.0944;%%

\end{thebibliography}
% example of Name_Of_Your_BibFile.bib
% @Article{Turcato:2006ch,
%      author    = "Turcato, M.",
%  collaboration = "ZEUS and H1",
%      title     = "Lepton flavour violation and charmonium physics at HERA",
%      journal   = "Nucl. Phys. Proc. Suppl.",
%      volume    = "162",
%      year      = "2006", 
%      pages     = "283-287",
%      SLACcitation  = "%%CITATION = NUPHZ,162,283;%%"
% }
% 
% @Unpublished{Gogitidze:2007du,
%      author    = "Gogitidze, N.",
%  collaboration = "H1", 
%      title     = "Prompt photons and particle momentum distributions at
%                   HERA", 
%      year      = "2007",
%      note    = "hep-ex/0701033",
%      SLACcitation  = "%%CITATION = HEP-EX 0701033;%%"
% }

\end{footnotesize}

% ****************************************************************************
% END OF BIBLIOGRAPHY AREA
% ****************************************************************************

\end{document}